\documentstyle[pre,aps]{revtex}

\newcommand{\dgr}{\dagger}
\newcommand{\rar}{\stackrel{\rightarrow}{r}}
\newcommand{\Rar}{\stackrel{\rightarrow}{R}}
\newcommand{\vt}{\vartheta}
\newcommand{\vp}{\varphi}
\newcommand{\ra}{\rightarrow}
\newcommand{\be}{\begin{equation}}
\newcommand{\ee}{\end{equation}}
\newcommand{\al}{\alpha}
\newcommand{\om}{\omega}
\newcommand{\bt}{\beta}
\newcommand{\ga}{\gamma}
\newcommand{\la}{\lambda}
\newcommand{\lt}{\left (}
\newcommand{\rt}{\right )}
\newcommand{\ep}{\varepsilon}

\begin{document}

\draft

\title{Evaporation and Condensation of Clusters}
\author{V. I. Yukalov and E. P. Yukalova} 
\address{Department of Mathematics and Statistics \\
Queen's University, Kingston, Ontario K7L 3N6, Canada \\
and \\
Bogolubov Laboratory of Theoretical Physics \\
Joint Institute for Nuclear Research, Dubna 141980, Russia}

\maketitle

\begin{abstract}

Influence of surrounding matter on the properties of clusters is considered 
by an approach combining the methods of statistical and quantum mechanics. 
A cluster is treated as a bound $\;N\;$--particle system and surrounding 
matter as thermostat. It is shown that, despite arbitrary strong interactions 
between particles, cluster energy can be calculated by using the controlled 
perturbation theory. The accuracy of the latter is found to be much higher 
than that of the quasiclassical approximation. Spectral distribution is 
obtained by minimizing conditional entropy. Increasing the thermostat 
temperature leads to the depletion of bound states. The characteristic 
temperature when bound states become essentially depleated defines the 
temperature of cluster evaporation. The inverse process of lowering the 
thermostate temperature, yielding the filling of bound states, corresponds 
to cluster condensation.
\end{abstract}

\section{Introduction}

The formation of clusters is a common tendency of matter. Clusters of quarks 
form nucleons, clusters of nucleons form nuclei, clusters of nuclei form 
atoms and molecules, clusters of the latter form various objects on the 
Earth, including alive beings. The Earth itself, as well as other planets, 
are clusters.  Planets clusterize around the Sun or around other stars. 
Galaxes in the Universe are also a kind of clusters.

Generally, a cluster is a compact group of objects bound together for 
sufficiently long time. More exactly, the lifetime of a cluster is to be 
much longer than the characteristic times describing its motion as a whole. 
In the ideal case, when the lifetime of a cluster is infinite, the latter is 
absolutely stable. Usually, clusters have finite lifetimes, thus, being 
metastable, although the lifetime can be extremely long.

In what follows, for concreteness, we shall mean under a cluster a group 
of $\;N\;$ particles forming a bound state because of attracting forces 
between them. There are two main approaches of treating clusters 
theoretically, that of statistical and of quantum mechanics. The statistical 
approach permits to define the relative concentrations of different clusters 
and other thermodynamic functions of a statistical systems, provided that the 
basic characteristics of each type of clusters, such as mass, binding energy, 
and quantum numbers, like spin, are given a priori. This approach describes, 
for example, the relative abundances of chemical elements in the universe 
[1]. In the droplet models of condensation [1,2] the noninteracting droplets 
are characterized by their volumes and surface tension. Alternatively, the 
interaction potentials between droplets are to be postulated [3]. The 
appearance of droplets is typical of the liquid--vapour phase transition 
occurring either in usual gases or in nuclear matter [4-6].

Recently, a great attention has been attracted to the use of the statistical 
approach for describing the deconfinement phase transition in quantum 
chromodynamics. When increasing temperature or density, hadron matter is 
expected to decompose into quark--gluon plasma [7-12]. High density of 
nucleons can be reached in ultrarelativistic nuclear collisions [13-16] and 
both high density and temperature, in heavy--ion collisions [17-20]. Numerical
simulations with lattice models [21-24] undoubtfully show that nonperturbative
phenomena dominate in a wide range around the deconfinement phase transition, 
up to temperatures of about $\;2T_c\;$ [25,26]. These nonperturbative 
phenomena are thought to be caused by the formation of quark and gluon 
clusters [27-32] inside quark--gluon plasma, similar to the appearance of 
liquid droplets in fog. The deconfinement--confinement phase transition, to 
our mind, is an analog of the evaporation--condensation phase transition.

In the quantum--mechanical approach, one considers the properties of an 
individual cluster in the vacuum and the interactions of a small number of 
clusters. For instance, in last years the properties of multiquark clusters 
have been intensively studied [33-38]. Such clusters, as various scattering 
experiments indicate [39-43], can exist in nuclei and even in deuteron 
[44,45]. More information on multiquark clusters can be found in reviews 
[46,47]. Other examples of clusters whose description is mainly based 
on quantum mechanics are nuclei themselves, alpha clusters [48], resonant 
nuclear clusters [49,50] and metallic clusters [51].

The properties of a cluster, defined when the latter is in the vacuum, can be
strongly changed if the cluster is inserted into matter. Moreover, the 
stability of a cluster can be essentially influenced by surrounding medium. 
Therefore, when one calculates the concentrations of clusters in a statistical
system, fixing cluster characteristics a priori, while the latter can be 
noticeably disturbed by the influence of matter, would yield wrong 
conclusions. At the same time, for understanding how matter influences cluster
properties, one needs to have a method of their accurate definition. To 
describe the properties of a cluster is usually not very easy since one has 
to deal with bound states of strongly interacting particles.

In the present paper we consider the following mutually related questions: (i)
How to accurately calculate the energy spectrum of bound states for an 
$\;N\;$--particle cluster? (ii) What are the statistical weights of such 
states? (iii) How the evaporation or condensation of a cluster is caused by 
surrounding matter?

The questions addressed above are, of course, too complicated to be 
exhaustively answered in one paper. Here, we suggest some principal ways of 
solving the problem and illustrate the ideas by a simplified model: A cluster 
consists of $\;N\;$ particles interacting with each other through attractive 
pair forces of power law; surrounding matter is replaced by thermostat. We 
hope that the suggested approach can be applied to more realistic problems. 
But first we have to be convinced that there exists the meth

od permitting to answer the above formulated questions in principle.

\section{Cluster energy}

Consider a system of $\;N\;$ particles defined by the Hamiltonian
$$ H_N =\int\psi^\dgr(\rar )K(\rar )\psi(\rar )d\rar + $$
\be
+\frac{1}{2}\int\psi^\dgr (\rar )\psi^\dgr (\rar ')\Phi(|\rar -\rar '|)
\psi(\rar ')\psi(\rar )d\rar d\rar ' ,
\ee
in which $\;K(\rar )\;$ is a kinetic--energy operator; $\;\psi(\rar )\;$ is a 
field operator being a column $\;[\psi_s(\rar )]\;$ in the space of internal 
degrees of freedom, such as spin, isospin and so on. The interaction potential
does not depend on internal degrees of freedom.

The average energy of the system can be exactly expressed through the 
second--order density matrix. There are several forms of such expressions 
[52-54]. In the second--quantization representation, which we use here, the 
average energy can be written as
$$ \langle H_N\rangle =\frac{1}{2(N-1)}\int\delta(\rar_1 -\rar_1')\delta 
(\rar_2  -\rar_2')H_2(\rar_1,\rar_2)\times $$
\be
\times\rho_2(\rar_1,\rar_2,\rar_1',\rar_2')d\rar_1d\rar_2d\rar_1'd\rar_2' ,
\ee
where the reduced Hamiltonian is
\be
H_2(\rar_1,\rar_2) \equiv K(\rar_1)+K(\rar_2) +(N-1)\Phi(|\rar_1-\rar_2|)
\ee
and the second--order density matrix
\be
\rho_2(\rar_1,\rar_2,\rar_1',\rar_2') =\langle \psi^\dgr(\rar_2')
\psi^\dgr(\rar_1')\psi(\rar_1)\psi(\rar_2)\rangle .
\ee
The form (2) can be easily obtained using the fact that, when the number of 
particles is fixed, the second-- and first--order density matrices are 
connected by the relation
$$ \int\rho_2(\rar_1,\rar_2,\rar_1',\rar_2)d\rar_2 =
(N-1)\rho_1(\rar_1,\rar_1') , $$
$$ \rho_1(\rar,\rar') =\langle\psi^\dgr(\rar')\psi(\rar)\rangle . $$

Let us note that there is another way, not mentioned in literature, of 
deriving (2). Write the Hamiltonian in the Schr\"odinger representation
\be
H(\rar_1,\rar_2,\ldots,\rar_N) =\sum_{i=1}^{N}K(\rar_i) +
\frac{1}{2}\sum_{i\neq j}^{N}\Phi(|\rar_i -\rar_j|) .
\ee
Passing to the Heisenberg representation, one transforms the first and second 
terms in (5) to the first and second terms in (1), respectively. However, 
nothing prohibits us to rewrite (5) in the identical form
$$ H(\rar_1,\rar_2,\ldots ,\rar_N) =
\frac{1}{2(N-1)}\sum_{i\neq j}^{N}H_2(\rar_i,\rar_j) , $$
in which $\;H_2\;$ is given by (3). From the latter expression, in the 
Heisenberg representation we obtain
\be
H_N =\frac{1}{2(N-1)} \int\psi^\dgr(\rar)\psi^\dgr(\rar')
H_2(\rar,\rar')\psi(\rar')\psi(\rar)d\rar d\rar' .
\ee
And from (6) one immediately gets (2).

The second--order density matrix (4) can be presented as an expansion
\be
\rho_2(\rar_1,\rar_2,\rar_1',\rar_2') = \sum_{\al\bt}B_{\al\bt}
\Psi_\al(\rar_1,\rar_2)\Psi_\bt^*(\rar_1',\rar_2')
\ee
over an orthonormalized basis $\;\{\Psi_\al\}\;$. The expansion coefficients, 
because of the normalization condition
$$ \int\rho_2(\rar_1,\rar_2,\rar_1,\rar_2)d\rar_1d\rar_2 = N(N-1) , $$
obey the sum rule
\be
\sum_{\al}B_{\al\al} = N(N-1) .
\ee
With the notation
$$ H_{\al\bt} \equiv \int\Psi_\al^*(\rar_1,\rar_2)H_2(\rar_1,\rar_2)
\Psi_\bt(\rar_1,\rar_2)d\rar_1 d\rar_2 , $$
the average energy (2) becomes
\be
\langle H_N\rangle =\frac{1}{2(N-1)}\sum_{\al\bt}B_{\al\bt}H_{\bt\al} .
\ee

It is natural to choose the basis $\;\{\Psi_\al\}\;$ in the expansion (7) as 
a set of eigenfunctions of the reduced Hamiltonian (3),
\be
H_2(\rar_1,\rar_2)\Psi_\al(\rar_1,\rar_2)  = E_\al\Psi_\al(\rar_1,\rar_2) .
\ee
Equation (10) describes quantum mechanics of a pair of particles. Such a pair 
can be called [54] a pairon. Defining
\be
p_\al \equiv \frac{B_\al}{N(N-1)} ; \qquad \sum_\al p_\al = 1 ,
\ee 
we see that $\;p_\al\;$ has the meaning of the probability that a pairon is 
in a state $\;\al\;$. Now, the average energy (9) reduces to
\be
\langle H_N\rangle =\frac{N}{2}\sum_\al p_\al E_\al .
\ee

For concreteness, let us take the kinetic--energy operator in the standard 
form
$$ K(\rar ) = -\frac{\nabla^2}{2m_0} , $$
$\;m_0\;$ being a particle mass. Then, using the center--of--mass and relative
coordinates
$$ \Rar \equiv \frac{\rar_1 +\rar_2}{2} , \qquad 
\rar \equiv \rar_1 -\rar_2 , $$
and introducing the masses
$$ M \equiv 2m_0, \qquad \mu \equiv \frac{m_0}{2} , $$
one can, as is known, factorize the eigenfunction in (10) into the product
$$ \Psi_\al(\rar_1,\rar_2) = 
\Psi_Q(\Rar)\frac{\chi_{nl}(r)}{r}Y_{lm}(\vt,\vp) , $$
in which $\;\Psi_q(\Rar)\;$ is a plane wave, with a wave vector 
$\;\stackrel{\ra}{Q}\;$, corresponding to a pairon center--of--mass motion; 
$\;Y_{lm}\;$ is a spherical harmonic; and the pairon relative motion is 
described by a function $\;\chi_{nl}\;$ satisfying the equation
\be
H_N(r)\chi_{nl}(\rar) = E_{nl}(N)\chi_{nl}(r)
\ee
with the relative Hamiltonian
\be
H_N(r) = -\frac{1}{2\mu}\frac{d^2}{dr^2} +\frac{l(l+1)}{2\mu r^2} +
(N-1)\Phi(r).
\ee
The pairon energy $\;E_\al\;$ in (10) reads
\be
E_\al =\frac{\stackrel{\ra}{Q}^2}{2M} + E_{nl}(N) .
\ee
Thus, the pairon quantum index $\;\al\;$ is specified as a set 
$\;\al =\{ \stackrel{\ra}{Q},n,l,m\}\;$ of six indices: three components in 
the wave vector $\;\stackrel{\ra}{Q}=\{ Q_a\; |a=1,2,3\} \;$, each 
$\;Q_a\in {\bf R}\;$; the radial quantum number $\;n\;$, which is 
$\;n=0,1,2,\ldots \;$ for bound states and $\; n\in {\bf R}\;$ for unbound 
states; the orbital quantum number $\;l=0,1,2,\ldots\;$ and the azimuthal 
quantum number $\;m=-l,-l+1,\ldots ,-1,0,1,\ldots ,l-1,l\;$.

Now we need to define the statistical weights of the energy levels (15). We 
assume that the considered system of $\;N\;$ particles is immersed into 
thermostat with temperature $\;T\;$. To find the weights $\;p_\al\;$ in (12) 
we can proceed as follows. Define the entropy
$$ S = -\sum_\al p_\al\ln p_\al . $$
The weights $\;p_\al\;$ have to satisfy the normalization condition in (11) 
and give an average energy per particle according to (12), that is
$$ \sum_\al p_\al = 1 , \qquad \frac{1}{N}\langle H_N\rangle =
\sum_\al p_\al E_\al . $$
In addition, there can exist some constraints or selection rules which usually
can be related to operators $\;\hat I\;$ being integrals of motion whose 
averages can be written as 
$$ \langle\hat I\rangle = \sum_\al p_\al I_\al . $$
For instance, $\;\hat I\;$ can be a projector to a space of a given symmetry. 
Introduce the notation
$$ \pi_\al \equiv\exp (-\ga I_\al) , $$
where $\;\ga\;$ is a Lagrange multiplier. Minimizing the conditional entropy
$$ \stackrel{-}{S} = S +\bt\lt\frac{1}{N}\langle H_N\rangle - 
\sum_\al p_\al E_\al \rt + $$
\be
+ \lt 1-\ln Z\rt\lt\sum_\al p_\al -1\rt +\ga \lt\langle \hat I
\rangle -\sum_\al p_\al I_\al \rt , 
\ee
with the Lagrange multipliers $\;\bt =T^{-1},\;\ga\;$, and $\;1-\ln Z\;$, we 
get
\be
p_\al =\frac{\pi_\al}{Z}\exp\lt-\bt E_\al\rt ; \qquad 
Z = \sum_\al \pi_\al\exp\lt-\bt E_\al\rt .
\ee

As an example of selection rules defining $\;\pi_\al\;$ consider the case 
when the second--order density matrix possesses the symmetry property
$$ \rho_2(\rar_2,\rar_1,\rar_1',\rar_2') = 
\pm\rho_2(\rar_1,\rar_2,\rar_1',\rar_2') . $$
Respectively, the functions $\;\Psi_\al\;$ in the expansion (7) are either 
symmetric or antisymmetric with respect to coordinate permutation,
$$ \Psi_\al(\rar_2,\rar_1) =\pm\Psi_\al(\rar_1,\rar_2) . $$
The permutation $\;\rar_1 \leftrightarrow\rar_2\;$ is equivalent to the 
transformation
$$ \Rar\ra\Rar , \qquad \rar\ra -\rar , \qquad \vt\ra\pi -\vt , \qquad 
\vp \ra \vp +\pi . $$
Using the property of spherical functions
$$ Y_{lm}\lt\pi-\vt ,\vp+\pi\rt = (-1)^lY_{lm}(\vt ,\vp ) , $$
we have
$$ \Psi_\al(\rar_2,\rar_1) =(-1)^l\Psi_\al(\rar_1,\rar_2) . $$
Therefore the coefficients $\;B_{\al\bt}\;$ in (7) should have the form
$$ B_{\al\bt} =\pi^\pm_\al\pi_\bt^\pm C_{\al\bt} , $$
in which
\begin{eqnarray}
\pi_\al^+\equiv\pi_l^+ =\left\{ \begin{array}{cc}
1; & l=0,2,4\ldots \\ \nonumber
0; & l=1,3,5,\ldots \end{array}
\right.
\end{eqnarray}
\begin{eqnarray}
\pi_\al^-\equiv\pi_l^- =\left\{ \begin{array}{cc}
0; & l=0,2,4\ldots \\ \nonumber
1; & l=1,3,5,\ldots \end{array}
\right.
\end{eqnarray}
In this case, $\;\pi_\al^\pm\;$ is a projector to even or odd states, 
respectively. Then, from (17) we derive
$$ p_{nl} \equiv\sum_{Q_m} p_{Qnlm} =\pi_l^\pm \frac{2l+1}{Z_{int}} 
\exp\left \{ -\frac{\bt}{2}E_{nl}(N)\right \} ; $$
\be
Z_{int} \equiv \sum_{nl}\pi_l^\pm(2l+1)\exp\left \{ -\frac{\bt}{2}E_{nl}(N) 
\right\} .
\ee

In this way, the average energy (12) takes the form
\be
\langle H_N\rangle = \frac{3}{2}NT + E(N) ,
\ee
where
\be
E(N) \equiv \frac{N}{2}\sum_{nl} p_{nl}E_{nl}(N) .
\ee
The interpretation of the obtained result is straightforward. Since we are 
thinking of the considered system as of an $\;N\;$--particle cluster, we may 
say that the first term in (19) is the kinetic energy of the cluster, and (20)
is the internal energy of particles inside the cluster. A part of the spectrum
$\;E_{nl}(N)\;$ must correspond to bound states. In order that we could, in 
principle, speak about an $\;N\;$--particle cluster, the number of bound 
states must be not less than $\:N\;$.

\section{Bound states} 

To be able to form bound states, the particles should interact with each 
other through forces containing an attractive part. The interaction potential 
in the reduced Hamiltonian (3) includes the factor $\;(N-1)\;$. When 
$\;N\gg 1\;$, then, even if the potential$\;\Phi\;$ is small, the product 
$\;(N-1)\Phi\;$ can be large. Moreover, the interaction between particles 
forming bound states is practically never small. All this means that the 
standard perturbation theory will not work for describing bound states. 
However, the problem is treatable by the controlled perturbation theory 
[55-58] which is applicable to arbitrary strong interactions.

For demonstrational purpose, let us take an attractive potential of power law
\be
\Phi(r) = -Ar^{-\nu} ,
\ee
in which $\;A,\;\nu>0\;$ and $\;r\in {\bf R_+}\;$. Then the Hamiltonian (14) 
reads
\be
H_N(r) = -\frac{1}{2\mu}\frac{d^2}{dr^2} + \frac{l(l+1)}{2\mu r^2} -
\lt N-1\rt\frac{A}{r^\nu} .
\ee
The latter, using a scaling, can be reduced to the dimensionless form
\be
H = -\frac{1}{2}\frac{d^2}{dr^2} +\frac{l(l+1)}{2r^2} -\frac{g}{r^\nu} .
\ee
The return from (23) to (22) is made by substituting
$$ H \ra \frac{H_N(r)}{\om}; \qquad 
\om \equiv\lt\mu^\nu A^2\rt^{1/(2-\nu)}, $$
\be
r\ra\sqrt{\mu\om}r , \qquad g \ra N-1 .
\ee 

As zero approximation we may accept the Hamiltonian
\be
H_0 = -\frac{1}{2}\frac{d^2}{dr^2} +\frac{l(l+1)}{2r^2} -\frac{u}{r}
\ee
with the Coulomb interaction including yet unknown parameter $\;u\;$. The 
eigenfunctions of (25) are
$$ \chi_{nl}^{(0)}(r) = \left [\frac{n!u}{(n+2l+1)!}\right ]^{1/2} 
\frac{1}{n+l+1}\lt\frac{2ur}{n+l+1}\rt^{l+1}\times $$
$$ \times \exp\lt -\frac{ur}{n+l+1}\rt L_n^{2l+1}\lt\frac{2ur}{n+l+1}\rt , $$
where $\;L_n^l\;$ is an associate Laguerre polynomial.
The eigenvalues of (25) are
\be
E_0(n,l,g,u) = -\frac{u^2}{2(n+l+1)^2} .
\ee
Starting from (25), we can invoke the Rayleigh--Schr\"odinger perturbation 
theory with the perturbation
$$ \Delta H =\frac{u}{r}-\frac{g}{r^\nu} . $$
In doing this, we encounter the integral
\be
J_{ns}^l(\nu) \equiv 
\int_{0}^{t}t^{2l+2+\nu}e^{-t}L_n^{2l+1}(t)L_s^{2l+1}(t)dt.
\ee
The latter can be presented as a sum. For instance, if $\;s=n\;$, then
$$ J_{nn}^l(\nu) = \frac{(n+2l+1)!}{n!} \sum_{p=0}^{n}
\frac{(-1)^p\Gamma(p+2l+3+\nu)}{p!(n-p)!\Gamma(p+2l+2)} \vp_n(-p-1-\nu) , $$
where
$$ \vp_n(x) \equiv \frac{\Gamma(n+x)}{\Gamma(x)} ; \qquad 
\vp_0(x) \equiv 1 . $$
In particular,
$$ J_{nn}^l(-1) =\frac{(n+2l+1)!}{n!} , \qquad 
J_{00}^l(-\nu) =\Gamma(2l+3-\nu). $$
For the first approximation we have
\be
E_1(n,l,g,u)=E_0(n,l,g,u) +\frac{u^2}{(n+l+1)^2} -
\frac{gu^\nu}{(n+l+1)^2}A_{nl}(\nu) ;
\ee
here
$$ A_{nl} \equiv \frac{n!J_{nn}^l(-\nu)}{(n+2l+1)!}
\lt\frac{2}{n+l+1}\rt^{\nu -1} , $$
$$ A_{0l} =\frac{\Gamma(2l+3-\nu)}{(2l+1)!}\lt\frac{2}{l+1}\rt^{\nu -1} . $$
The procedure can be continued yielding $\;E_k(n,l,g,u)\;$. Then, at each step
$\;k\;$, the unknown parameter $\;u\;$ is to be transformed to a function 
$\;u_k(n,l,g)\;$, such that the sequence $\;\{e_k(n,l,g)\}_{k=0}^\infty\;$ 
of terms
\be
e_k(n,l,g) \equiv E_k(n,l,g,u_k(n,l,g))
\ee
would be convergent. Because of their role of forcing the convergence of the 
approximation sequence, the functions $\;u_k(n,l,g)\;$ are called control, or 
governing, functions. Their explicit form is defined by fixed--point 
conditions. One of the simplest such conditions is the condition of minimal 
sensitivity [59],
\be
\frac{\partial}{\partial u_k}E_k(n,l,g,u_k) = 0 .
\ee
When equation (30) has no solution for the control function $\;u_k\;$, one 
can use other types of fixed--point conditions [55-58]. If there exist several
solutions for $\;u_k\;$, than one has to resort to the stability analysis 
[57,60] choosing that solution which makes the procedure more stable.

In the first approximation we have only one solution for the control function
\be
u_1(n,l,g) =\left [\nu A_{nl}(\nu)g\right ]^{1/(2-\nu)} .
\ee
From (28) and (29) we find
\be
e_1(n,l,g) = -\frac{(2-\nu)u_1^2(n,l,g)}{2\nu(n+l+1)^2} .
\ee
Eq.(32) shows that for $\;\nu \geq 2\;$ there are no bound states since all 
energy levels become positive moving to the continuous spectrum. When 
$\;\nu =1\;$, then
$$ u_1(n,l,g) = g, \qquad A_{nl}(1) = 1 \quad (\nu =1) , $$
and (32) yields
$$ e_1(n,l,g) = -\frac{g^2}{2(n+l+1)^2} \qquad (\nu = 1) , $$
that is, we get an exact spectrum corresponding to the Coulomb potential.

For asymptotically small powers $\;\nu\ra 0\;$, since
$$ J_{nn}^l(0) =2(n+l+1)\frac{(n+2l+1)!}{n!} , \qquad A_{nl}(0)=(n+l+1)^2 , $$
from (31) we have
$$ u_1(n,l,g) \simeq (n+l+1)\sqrt{g\nu} \qquad (\nu\ra 0) . $$
Therefore (32) gives
$$ \lim_{\nu\ra 0}e_1(n,l,g) = -g . $$

In the case of $\;n=0\;$, (32) becomes
\be
e_1(0,l,g) = -\frac{2-\nu}{8\nu}
\left [ \frac{2\nu\Gamma (2l+3-\nu)}{(l+1)(2l+1)!}g\right ]^{2/(2-\nu)} .
\ee
And the ground--state energy level, when $\;n=l=0\;$, is
\be
e_1(0,0,g) = -\frac{2-\nu}{8\nu}\left [ 2\nu\Gamma(3-\nu)g
\right ]^{2/(2-\nu)} .
\ee

We calculated also the second--order approximation $\;e_2(n,l,g)\;$ and the 
corrected first approximation, following the method of refs.[57,58,60]. The 
latter approximation has the form
\be
e_1^*(n,l,g) = e_1(n,l,g)\exp\{\lambda_1(n,l)\delta_*\} ,
\ee
in which $\;\lambda_1(n,l)\;$ is a local Lyapunov exponent for the 
approximation flow and $\;\delta_*\;$ is a damping parameter. However, we 
shall limit ourselves by the first approximation (32) because the expressions 
for the second and for the corrected first approximations are essentially more
cumbersome, althougt, as numerical analysis has shown, they do not  differ 
much in accuracy from the first approximation.

The stability of the calculated procedure in the controlled perturbation 
theory [55-58,60] can be checked in the following way. Define the coupling 
function $\;g_k(n,l,f)\;$ by the equation
\be
E_0(n,l,g,u_k(n,l,g)) = f .
\ee
This, in the case of (26) and (31), leads to the equality
$$ u_1(n,l,g) =(n+l+1)\sqrt{-2f} . $$
For the coupling function we obtain
\be
g_1(n,l,f) =\frac{(n+l+1)^{2-\nu}}{\nu A_{nl}(\nu)}\lt -2f\rt^{1-\nu/2} .
\ee
Now we have to define the points of the trajectory of the approximation 
cascade as
\be
y_k(n,l,f) \equiv e_k(n,l,g_k(n,l,f)) .
\ee
The starting point is $\;y_0(n,l,f) = f\;$ and the first point is
\be
y_1(n,l,f) =\frac{2-\nu}{\nu} f.
\ee
The mapping multipliers are
\be
\mu_k(n,l,f) \equiv \frac{\partial}{\partial f} y_k(n,l,f) ,
\ee
so that
\be
\mu_1(n,l,f) =\frac{2-\nu}{\nu} .
\ee
A necessary stability condition is
\be
|\mu_k(n,l,f)| < 1 ,
\ee
which shows the existence of contraction at the $\;k\;$--step with respect to 
the starting point. As we see, (41) satisfies (42) if $\;\nu > 1\;$.

Continuing the stability analysis for the second point of the cascade 
trajectory, we come to the mapping multiplier
\be
\mu_2(n,l,f) =\mu_1(n,l,f) +\la_1(n,l) .
\ee
Numerical calculations evince that the maximal Lyapunov exponent
\be
\la_1 \equiv\sup_{n,l}\la_1(n,l)
\ee
is positive and small, $\;0<\la_1\ll 1\;$. Whence, the maximal mapping 
multiplier (43) is
$$ \sup_{n,l}|\mu_2(n,l,f)| = \left |\frac{2-\nu}{\nu} +\la_1\right | . $$
The stability condition (42) for the mapping multiplier (43) is uniformly 
valid for all $\;n,l\;$ if 
\be
\nu > \frac{2}{2-\la_1} \simeq 1 +\frac{\la_1}{2} .
\ee
Another stability condition requires that the Lyapunov exponents 
$\;\la_k\;$ would be negative [58,60].

To check the accuracy of the procedure, let us introduce the notation
\be
e_k(n,l) \equiv e_k(n,l,1),
\ee
 using which (29) can be written as
\be
e_k(n,l,g) = e_k(n,l)g^{2/(2-\nu)} .
\ee
The energy levels (46) can be compared with those obtained by direct numerical
solution of the corresponding Schr\"odinger equation [61,62].

The Schr\"odinger equation with the Hamiltonian (23) can also be solved in 
the quasiclassical approximation [63] which yields
$$ e_{WKB}(n,l,g) = -\left \{ \sqrt{\frac{2}{\pi}} 
\frac{2-\nu}{2(2-\nu)(n+l+1)+(\nu -1)(2l+1)} \times \right. $$
\be
\left. \times \Gamma\lt\frac{2-\nu}{2\nu}\rt {\Bigl /}
\Gamma\lt\frac{1}{\nu}\rt \right \}^{2\nu/(2-\nu)}g^{2/(2-\nu)} . 
\ee
There exists a generalization of the quasiclassical approximation based on 
the use of modified Airy functions [64], but this technique, as it seems, is 
useful only for one--dimensional problems.

Introducing the notation
\be
e_{WKB}(n,l) \equiv e_{WKB}(n,l,1) ,
\ee
we can cast (48) into
\be
e_{WKB}(n,l,g) = e_{WKB}(n,l) g^{2/(2-\nu)} .
\ee
The form (49) is convenient to compare with direct numerical solutions [61,62].

The most proper characteristic of the precision of an approximate method is, 
apparently, the relative error. Thus, the error of (46) is defined as a 
percentage deviation
$$ \ep_k(n,l) \equiv \frac{e_k(n,l)-e(n,l)}{|e(n,l)|}\cdot 100\% $$
from an exact numerical value $\;e(n,l)\;$. The percentage errors of (35) and 
(48) are defined in the same way, as $\;\ep_k^*(n,l)\;$ and 
$\;\ep_{WKB}(n,l)\;$, respectively. The uniform accuracy of an approximation, 
with respect to all energy levels, is characterized by the maximal error
$$ \ep_k \equiv \sup_{n,l}|\ep_k(n,l)| . $$
Analogously, the maximal errors $\;\ep_k^*\;$ and $\;\ep_{WKB}\;$ are defined.

The accuracy of the controlled perturbation theory and that of the 
quasiclassical approximation, for power--law potentials, is demonstrated in 
tables I--III, as compared to the exact numerical values $\;e(n,l)\;$  and 
with each other. In the table I the results are presented for the potential 
(21) with the power $\;\nu =0.2\;$. The maximal errors are $\;\ep_1 =0.8\%\;$ 
and $\;\ep_{WKB}=4\%\;$. The errors for the second approximation 
$\;e_2(n,l)\;$ and for the corrected first approximation $\;e_1^*(n,l)\;$ are 
not displyed since they are very close to $\;e_1(n,l)\;$. The accuracy of (35)
in the case of the damping parameter $\;\delta_*=1\;$ is practically the same 
as in the case of $\;\delta_*=1/2\;$, for the latter case being slightly 
better. The accuracy, when passing from $\;e_1(n,l)\;$ to $\;e_2(n,l)\;$, 
does not improve because for $\;\nu=0.2\;$ the stability condition (45) is 
not valid. The corrected approximation $\;e_1^*(n,l)\;$ is not more accurate 
than $\;e_1(n,l)\;$ since the Lyapunov exponent (44) is positive.

Table II presents the calculations for $\;\nu=0.8\;$. The maximal errors are 
$\;\ep_1=0.3\%\;$ and $\;\ep_{WKB} =6\%\;$. The accuracy of $\;e_2\;$ and 
$\;e_1^*\;$ is pretty much the same as of $\;e_1\;$. The maximal error 
$\;\ep_1^*\;$ practically does not change when replacing $\;\delta_*=1\;$ by 
$\;\delta_* =1/2\;$. The absence of the improvement in accuracy is again 
caused by the fact that the stability conditions do not hold for 
$\;\nu=0.8\;$. Nevertheless, as in the previous case of $\;\nu=0.2\;$, the 
controlled perturbation theory is much more accurate than the quasiclassical 
approximation.

Table III contains the results for $\;\nu \geq 1\;$. For $\;\nu=1\;$, all 
approximations become exact answers. For $\;\nu >1\;$ the stability condition 
(45) now holds, and the accuracy of $\;e_2\;$ is slightly higher than of 
$\;e_1\;$. The Lyapunov exponent (44), as earlier, is positive, so 
$\;e_1^*\;$ does not improve $\;e_1\;$. Again, the quasiclassical 
approximation is essentially less precise than the controlled perturbation 
theory.   

\section{Evaporation temperature}

For the considered case of the power--law attraction (21), the spectrum of 
the relative motion of particles inside a cluster, defined by eq.(13), has 
the form
\be
E_{nl}(N) =\om e(n,l)g^{2/(2-\nu)} ,
\ee
where
$$ \om =(\mu^\nu A^2)^{1/(2-\nu)} , \qquad \mu =\frac{m_0}{2}, 
\quad g =N-1 . $$
The value of $\;e(n,l)\;$, as is shown in the previous section, can be 
calculated, with a good accuracy, by the controlled perturbation theory 
[55-58]. For example, we can use $\;e_1(n,l)\;$ from (32).

Define the ground--state level of (51) as
\be
E_0(N) \equiv E_{00}(N) \qquad (n=l=0) .
\ee
Here we should substitute
\be
e_1(0,0) =-\frac{2-\nu}{8\nu} \left [ 2\nu 
\Gamma (3-\nu)\right ]^{2/(2-\nu)} .
\ee

The internal energy of particles inside a cluster is given by (20) with the 
spectral weights (18). The whole spectrum has two parts, discrete and 
continuous. The discrete spectrum corresponds to bound states with negative 
energies (51). The continuous spectrum is related to unbound states with 
positive energies.

When temperature increases from zero, the internal energy of a cluster (20) 
increases from a negative value to zero and then to positive values. The 
characteristic temperature, $\;T_c\;$, at which the cluster ceases to exist, 
can be defined as the temperature where $\;E(N)\;$ becomes zero,
$$ E(N)=0 \qquad (T = T_c) . $$
This temperature may be named the evaporation temperature. The absolute 
value $\;|E(N)|\;$ is called the binding energy. So, the evaporation 
temperature is that where the binding energy is zero.

Approximately, the evaporation temperature can be obtained as follows. Note 
that when the temperature of thermostat is such that $\;2T \ll |E_0(N)|\;$, 
then the lower bound states are mainly occupied, while the unbound states of 
continuous spectrum are practically empty. This situation describes a well 
defined cluster. Recall that the thermostat here models surrounding matter, 
and temperature can be interpreted as a measure for the kinetic energy of 
collisions. If the latter are intensive, so that $\;2T \gg |E_0(N)|\;$, then 
the bound states are strongly depleted and the unbound states are essentially 
populated. This means that the cluster itself is actually destroyed, since a 
cluster is well defined only if the particles are bound inside it. The 
characteristic temperature
\be
T_c(N) =\frac{1}{2}\left | E_0(N)\right | 
\ee
specifying the desintegration of a cluster can be called the evaporation 
temperature, or the condensation temperature if the inverse process of 
lowering temperature is considered. The energy $\;E_0(N)\;$ in (54), defined 
in (52), is the ground--state energy of a pair of particles from the 
collection of $\;N\;$ particles. This explains the appearance of the factor 
$\;\frac{1}{2}\;$ in (54).

It is convenient to introduce the dimensionless evaporation temperature
\be
t_c(N) \equiv \frac{T_c(N)}{\om} .
\ee
For the latter, taking account of (51) and (53), we have
\be
t_c(N) =\frac{2-\nu}{16\nu}\left [ 2\nu\Gamma(3-\nu)(N-1)
\right ]^{2/(2-\nu)} .
\ee
In the case of the purely Coulomb interaction $\;(\nu=1)\;$, (56) reduces to 
the expression
$$ t_c(N) =\frac{1}{4}\lt N-1\rt^2 \qquad (\nu = 1) . $$
Then (54) is recovered by multiplying the latter by $\;\om\;$ with
$$ \om =\frac{\al^2}{2}m_0c^2 ; \qquad \al \equiv \frac{A}{\hbar c} \quad 
(\nu =1) . $$
if $\;A =e^2\;$, where $\;e\;$ is the electron charge, then $\;\al =1/137\;$ 
is the fine--structure constant. In the latter case, $\;\om=13.613eV\;$, or 
$\;15.796\times 10^4K\;$.

The evaporation of a cluster is the manifestation of its instability against 
excitations caused by the influence of surrounding. There can exist another 
instability limiting the size of a cluster. This is related to the definition 
of the cluster mass
\be
M_N \equiv m_0N +\frac{E(N)}{c^2} \geq 0 
\ee
as of a nonnegative quantity. Here $\;E(N)\;$ is the internal energy of 
particles inside the cluster, being given by (20). Due to the dependence of 
$\;E(N)\;$ on temperature, the mass (57) is also dependent on $\;T\;$, so that
$\;M_N\;$ increases together with $\;T\;$. Let us introduce a function 
$\;\ga(T)\;$ by the relation
\be 
E(N) \equiv \frac{N}{2}\ga(T)E_0(N) ,
\ee
in which
$$ E_0(N) =\om e(0,0)(N-1)^{2/(2-\nu)} . $$
At zero temperature, $\;0<\ga(0)\leq 1\;$. With the notation (58) the cluster 
mass (57) reads
\be
M_N = m_0N + N\frac{\om\ga(T)}{2c^2}e(0,0)(N-1)^{2/(2-\nu)} .
\ee
As far as for $\;T<T_c\;$ the function $\;\ga(T)\;$ is positive, and 
$\;e(0,0)\;$ is always negative, the second term in (59) is negative. 
Therefore, with increasing $\;N\;$, the cluster mass (59) decreases and 
becomes zero at $\;N=N_0(T)\;$ which is 
\be
N_0(T) = 1 +\left [ -\frac{2m_0c^2}{\om\ga(T)e(0,0)}\right ]^{1-\nu/2} .
\ee
In the case of $\;\nu=1\;$, since $\;e(0,0)=-1/2\;$, we have
$$ N_0(T) = 1+\sqrt{\frac{8}{\al^2\ga(T)}} \quad (\nu =1) . $$
If we assume that $\;\al =1/137\;$ and take $\;\ga(T) \approx 1/2\;$, then 
$\;N_0\approx 549\;$.

Eq.(60) defines the maximal number of particles in a cluster. If $\;N\;$ 
surpasses $\;N_0\;$, the cluster desintegrates into several smaller clusters. 
With temperature increasing from zero to $\;T_c\;$, the maximal number of 
particles (60) increases. When $\;T\ra T_c -0\;$, then $\;\ga(T)\ra 0\;$. If 
we suppose that $\;\ga(T)\sim T - T_c\;$ as $\;T\ra T_c\;$, then 
$\;N_0(T)\sim (T-T_c)^{1-\nu/2}\;$. That is, the maximal number of particles 
in a cluster tends to infinity with the critical index $\;1-\nu/2\;$. The 
function $\;\ga(T)\;$ plays here the role of an order parameter.

The evaporation of a cluster, though reminds a phase transition, is, of 
course, not the same. Generally, we can consider a cluster of any number of 
particles $\;N>1\;$. Even we may take $\;N=2\;$. It would be too bold to 
speak about a phase transition for two particles. An actual 
evaporation--condensation phase transition occurs in an ensemble of many 
clusters, or droplets, which interact with surrounding as well as with each 
other. However, if the number of particles forming a cluster is very large, 
say $\;N\ra\infty\;$, then the evaporation of such a cluster can have many 
features of a phase transition. For example, a crystal can be treated as a 
cluster. The transition of the particles of this cluster from bound to unbound
states happens at the melting temperature $\;T_m\;$ which is an analog of the 
evaporation temperature. The temperature $\;T_m\;$ can be estimated as follows
[65]. According to the meaning of the transition from bound to unbound states,
the related characteristic temperature is to be close to the ground--state 
binding energy, $\;T_m\approx |E_0|\;$. The latter is about the effective 
depth of the interaction potential $\;|E_0|\approx \Phi_m\;$. So,
$$ T_m =C\Phi_m , $$
where $\;C\;$ is a constant of the order of one. Take, e.g., the elements of 
the inert group interacting through the $\;12-6\;$ Lennard--Jones potential. 
Then $\;\Phi_m=\ep\;$, where $\;\ep\;$ is a constant of the Lennard--Jones 
potential. Estimates give $\;C=1/\sqrt{2}\;$. The formula
$$ T_m =\frac{\ep}{\sqrt{2}} $$
evaluates the melting temperatures surprisingly well, which is demonstrated 
in table IV for $\;Ne,\; Ar,\; Kr\;$ and $\;Xe\;$ at zero pressure: The 
calculated $\;T_m\;$ is very close to the experimental value $\;T_m^{exp}\;$.

\section{Conclusion}

A cluster is defined as an object by $\;N\;$ particles in bound states. For 
any, arbitrary strong, interactions between particles the spectrum of bound 
states can be calculated with a good accuracy by means of the controlled 
perturbation theory. The latter is about an order more accurate than the 
quasiclassical approximation.

The properties of a cluster in the vaccuum and inside matter can be 
essentially different. The differences are traced by analysing a simple model 
when surrounding matter is replaced by thermostat. With increasing 
temperature, the cluster begins to evaporate. At the characteristic 
temperature of evaporation, $\;T_c\;$, the cluster desintegrates into its 
constituents. The evaporation temperature is approximately equal to one half 
of the ground--state binding energy.

The number of particles in a cluster is limited by $\;N_0\;$ depending on the 
interaction potential between particles and the thermostat temperature. The 
limit $\;N_0\;$ increases with temperature, tending to infinity as 
$\;T\ra T_c\;$.

The term evaporation is applicable to clusters of any number of particles 
$\;N>1\;$. The evaporation of small clusters is rather a continuous process 
than an abrupt transformation. The evaporation of large clusters can acquire 
features of phase transitions.

\vspace{5mm}

{\bf Acknowledgement} 

\vspace{3mm}

We are very grateful to Professor A.J.Coleman for many useful discussions and 
support.

\newpage

\begin{center}
{\bf Table Captions}
\end{center}

\vspace{5mm}

{\bf Table I.} The accuracy of the energy levels calculated using the 
controlled perturbation theory and the quasiclassical approximation, for the 
interaction potential with $\;\nu=0.2\;$.

\vspace{5mm}

{\bf Table II.} The same as in table I, but for the potential with the power 
$\;\nu=0.8\;$.

\vspace{5mm}

{\bf Table III.} The accuracy of the ground--state energy found by using the 
controlled perturbation theory and the quasiclassical approximation, for 
potentials with different powers.

\vspace{5mm}

{\bf Table IV.} The estimation of the melting temperature at zero pressure, 
for the inert--group crystals with the Lennard--Jones interaction.

\newpage

\begin{center}

{\bf Table I}

\vspace{0.5cm}

\begin{tabular}{|c|c|c|c|c|c|}\hline
$n$ & $l$ & $e(n,l)$ &$\ep_1(\%)$&$\ep_{WKB}(\%)$&$|\ep_{WKB}/\ep_1|$\\ \hline
0   &  0  & --0.72526 &   0.50     & --1.2          &    2.4    \\ 
    &  1  & --0.63319 &   0.29     & --2.8          &    9.7  \\ 
    &  2  & --0.58215 &   0.19     & --3.5          &    18 \\ \hline 
1   &  0  & --0.60834 & --0.32     & --0.40         &    1.3 \\
    &  1  & --0.56730 &   0.19     & --1.3          &    6.8 \\
    &  2  & --0.53733 &   0.24     & --2.0          &    8.3 \\ \hline
2   &  0  & --0.55191 & --0.76     & --0.22         &   0.29 \\
    &  1  & --0.52680 & --0.04     & --0.83         &     21 \\
    &  2  & --0.50628 &   0.14     & --1.4          &     10 \\ \hline
\end{tabular}

\vspace{1cm}

{\bf Table II}

\vspace{0.5cm}

\begin{tabular}{|c|c|c|c|c|c|}\hline
$n$ & $l$ & $e(n,l)$ &$\ep_1(\%)$&$\ep_{WKB}(\%)$&$|\ep_{WKB}/\ep_1|$\\ \hline
0   &  0  & --0.48336 &   0.21     & --2.0          &    9.5    \\ 
    &  1  & --0.19843 &   0.05     & --4.9          &    98  \\ 
    &  2  & --0.11707 &   0.20     & --5.8          &    29 \\ \hline 
1   &  0  & --0.18334 & --0.10     & --0.58         &    5.8 \\
    &  1  & --0.11151 &   0.29     & --2.2          &    7.6 \\
    &  2  & --0.07739 &   0.21     & --3.5          &    17 \\ \hline
2   &  0  & --0.10517 & --0.27     & --0.18         &   0.67 \\
    &  1  & --0.07421 & --0.08     & --1.5          &     19 \\
    &  2  & --0.05635 &   0.17     & --2.3          &     14 \\ \hline
\end{tabular}

\vspace{1cm}

{\bf Table III}

\vspace{0.5cm}

\begin{tabular}{|c|c|c|c|c|c|}\hline
$\nu$& $e(0,0)$ &$\ep_1(\%)$&$\ep_2(\%)$&$\ep_{WKB}(\%)$&$|\ep_{WKB}/\ep_2|$\\ \hline
1    & --0.5     &   0 &  0  &   0 &   1    \\ 
1.25 & --0.69938 & 1.4 & 1.2 & 7.1 & 5.9  \\ 
1.5  & --2.3687  & 12  & 11  & 28  & 2.5  \\ \hline 
\end{tabular}

\vspace{1cm}

{\bf Table IV}

\vspace{0.5cm}

\begin{tabular}{|c|c|c|c|} \hline
    &$\ep(K)$&$T_m=\frac{\ep}{\sqrt{2}}(K)$&$T_m^{exp}(K)$ \\ \hline
$Ne$& 36  & 25  & 24  \\
$Ar$& 121 & 86  & 84 \\
$Kr$& 163 & 115 & 117 \\
$Xe$& 232 & 163 & 161 \\ \hline
\end{tabular}

\end{center}
\end{document}